# The Impact of Trade and Financial Openness on Operational Efficiency and Growth: Evidence from Turkish Banks

Haibo Wang[1], Lutfu S. Sua[2], Burak Dolar[3]

[1]*Division of International Business and Technology Studies, A.R. Sánchez Jr. School of Business*, *Texas A&M International University, Laredo, TX, USA, hwang@tamiu.edu*
[2]*Department of Management and Marketing, Southern University and A&M College, Baton Rouge, LA, USA, lutfu.sagbansua@sus.edu*
[3]*Department of Accounting, Western Washington University, Bellingham, WA, USA, Burak.Dolar@wwu.edu*

**Abstract:**
This paper examines the relationship between trade and financial openness, as well as the operational efficiency and growth of Turkish banks, from 2010 to 2023. Utilizing CAMELG-DEA and dynamic panel data analysis, the study finds that increased trade openness significantly enhances banking efficiency, primarily due to heightened demand for banking services related to international trade. Financial openness further boosts growth by facilitating capital flows, expanding banks' credit portfolios, and increasing fee income from cross-border transactions. However, poverty levels have a negative impact on bank performance, reducing financial intermediation and innovation opportunities. The results underscore the crucial role of trade and financial openness in fostering banking sector growth in developing economies.

## 1. Introduction
### 1.1. Global Trade and Financial Openness Trends

Global trade openness, measured as the sum of exports and imports relative to the global GDP, has generally increased since the 1970s, rising from 25% in 1970 to 63% in 2022, according to the World Bank data. The mid-1980s marked the beginning of significant trade liberalization efforts, notably in developing countries, through structural economic and market reforms. Formation of regional trading blocs (such as the EU's Single Market in 1993 and NAFTA in 1994), the establishment of the World Trade Organization (WTO) in 1995, China's acceptance into the WTO in 2001, and the general acceleration of globalization played critical roles in enhancing trade openness. Additionally, advances in transportation and communication technologies, along with the digital revolution of the early 2000s, significantly lowered trade costs and facilitated the growth of more efficient global value chains, integrating emerging markets in Asia, Latin America, and Eastern Europe more deeply into the global economy. Consequently, global trade as a percentage of the world economy saw a rapid and steady increase from the mid-1980s until it peaked before the 2008 Global Financial Crisis. The period following the 2008 crisis generally displayed a declining trend, with trade openness plateauing at the 2008 peak, which was not again reached until 2022. (https://data.worldbank.org/indicator/NE.TRD.GNFS.ZS)

The global financial openness, measured by the sum of Net Foreign Assets (i.e., the value of a country's foreign assets minus the value of assets owned by foreigners within the country) and Net Incurrence of Liabilities (i.e., foreign financing obtained from foreigners minus the repayment of those liabilities) as a percentage of GDP, also showed a similarly rapid increase beginning in the early 1990s until the dawn of the Global Financial Crisis. During this period, advanced economies, along with emerging markets, saw a significant increase in financial openness, while less-developed countries experienced a more muted rise in openness. After 2008, financial openness generally stalled or declined worldwide, mirroring the trend in trade openness. The decline in global trade and financial openness after 2008 was influenced by several interrelated factors, including the global economic slowdown, structural shifts in the current financial order, rising geopolitical tensions, the rise in protectionism, the decline in foreign direct investment (FDI) volumes, and tighter financial regulations and capital controls aimed to protect economies from volatile capital flows. (https://econbrowser.com/archives/2016/07/chinn-ito-financial-openness-index-updated-to-2014)

### *1.2. Trade and Financial Openness Trends in Turkey*

Turkey's path to trade and financial openness, since the mid-1980s, has to a large extent followed global trends, particularly for emerging economies. Turkey was one of the first developing economies that embarked on a series of economic reforms aimed at liberalizing its economy and integrating the country into the newly emerging liberal global economic system. Economic reforms included export incentives, tariff reductions, greater capital inflows and outflows, and the lifting of controls on foreign exchange transactions. The signing of a customs union agreement with the European Union in 1996 played a significant role in integrating the Turkish economy with Europe and significantly boosted trade volumes with EU countries.

Turkey continued its economic liberalization and reforms into the 2000s, which made it an attractive destination for FDI. The period after 2010 has seen fluctuating trends in trade and financial openness influenced by various factors. Following the global financial crisis of 2008-2009, Turkey experienced a period of robust economic growth, driven by high domestic demand and increased exports. However, in the second half of the 2010s, Turkey entered a period of economic instability marked by high inflation, a widening current account deficit, and rising external debt, the effects of which were exacerbated by geopolitical tensions in the region. A severe currency crisis in 2018 led to a sharp depreciation of the Turkish lira, which negatively impacted financial stability and trade volumes. Following 2018, in an effort to stabilize the economy, address inflation, and manage external imbalances, policymakers generally continued to promote trade and financial openness, while capital controls have occasionally been used to stabilize the currency.
(World Bank, 2021, "Turkey Economic Monitor: Navigating the Waves.")

## 2. Theoretical Background and Hypotheses

This section provides a brief overview of theories in the literature as they relate to trade and financial openness.

### *2.1 Relevant Theories*

This section provides a brief overview of the relevant theories in the literature as outlined in Table 1.

**Table 1.** Relevant Theories

| Theory | Review |
|---|---|
| Financial intermediation theory | Allen and Santomero (1997) discuss the role of intermediation in the context of new markets for options and financial futures, focusing on intermediaries rather than companies or individuals, and emphasizing risk trading and participation costs. (Allen & Santomero, 1997; Baker, De Long, & Krugman, 2005; Eichenbaum, 1991; Ethier, 1975; McGrattan & Prescott, 2014; Scholtens & van Wensveen, 2000) Scholtens & Van Wensveen (2000) suggest how the financial intermediation theory can be further improved to better |

| | understand present-day phenomena in the financial sector (Scholtens & Van Wensveen, 2000). |
|---|---|
| Population and Growth Theory | Baker et al. (2005) discuss the correlation between the rates of return on assets and the rates of economic growth, arguing that the reduction in asset returns is greater for a given reduction in productivity growth (Baker, De Long, & Krugman, 2005). |
| Real Business-Cycle Theory | Eichenbaum (1991) assesses the empirical plausibility of the view that aggregate productivity stocks account for most of the variability in economic growth. Unlike other leading theories of the business cycle, the RBC theory sees business cycle fluctuations as the efficient response to exogenous changes in the real economic environment (Eichenbaum, 1991) |
| Monetary Growth Theory | In his study on financial assets and economic growth, Ethier (1975) discusses the role of inflation in a Keynesian economy (Ethier, 1975) |
| Basic Theory | Incorporating intangible investments into the Basic Theory, McGrattan & Prescott (2014) argue that measured labor productivity rises if the fall in output is underestimated, which occurs when there are large unmeasured intangible investments (McGrattan & Prescott, 2014). |

## 2.2 Trade and Financial Openness

Arif and Rawat (2019) analyze the impact of trade and financial sector openness on the financial development of South Asian countries using panel unit root and co-integration tests. The results show a significant negative effect of economic openness and a positive impact of trade openness on the financial development of the countries investigated (Arif & Rawat, 2019; Ashraf, Qian, & Shen, 2021; Baltagi, Demetriades, & Law, 2009; Bekaert, Harvey, & Lundblad, 2011; Khan, Hassan, Paltrinieri, & Bahoo, 2021; Kim, Lin, & Suen, 2010a, 2010b; Le, Kim, & Lee, 2016; Luo, Tanna, & De Vita, 2016; Niroomand, Hajilee, & Al Nasser, 2014; Rahman, Rahman, Rahman, & Masud, 2023; Zhang, Zhu, & Lu, 2015).

Based on a dataset involving 35 emerging economies, Ashraf et al. (2021) analyze the effect of trade and financial openness on bank loan pricing and report that higher levels of trade and financial openness lead to lower interest rates on gross bank loans, contributing to financial development in emerging economies (Ashraf, Qian, & Shen, 2021). Using dynamic panel estimation and data from developing and industrial countries, Baltagi et al. (2009) examine the relationship between trade, financial openness, and financial development, reporting that the marginal effects of trade openness have an inverse relationship with the level of financial openness (Baltagi, Demetriades, & Law, 2009).

Bekaert et al. (2011) employ panel regression to show that the impact of openness on factor productivity growth is more important than the effect on capital growth and claim that the growth boost from openness outweighs the detrimental loss in growth from global and regional banking crises (Bekaert, Harvey, & Lundblad, 2011). Using panel data for 87 countries, Kim et al. (2010a) investigate the short- and long-run relationships between financial development and trade openness and show that long-run complementarity between financial development and trade openness coexists with short-run substitutionary between the two policy variables (Kim, Lin, & Suen, 2010a, 2010b)

Focusing on financial development in the Gulf Cooperation Council countries, Khan et al. (2021) examine the comparative impact of trade and financial openness on Islamic and conventional banks, reporting a positive impact on the profitability of Islamic banks. The results further indicate that the positive marginal effect of trade openness on bank profitability decreases as the level of financial openness increases (Khan,

Hassan, Paltrinieri, & Bahoo, 2021). Applying the dynamic generalized methods of movements to a panel data set of 26 countries, Le et al. (2016) examine the determinants of financial sector development in Asia and Pacific and conclude that governance and institutional quality lead to development of financial sector in emerging economies while trade openness and economic growth are the main determinants of financial depth in such economies (Le, Kim, & Lee, 2016).

Luo et al. (2016) examine the relationship between financial openness, bank risk, and bank profit efficiency by utilizing commercial bank data from 140 countries. The results indicate that financial openness directly reduces bank profit efficiency while indirectly increasing bank risk through decreased profit efficiency (Luo, Tanna, & De Vita, 2016). Niroomand et al. (2014) examine the relationship between financial market development and trade openness by developing a long- and short-run model for 18 emerging economies and report that financial market development has a significant effect on trade openness in both the short and long run for the majority of the countries investigated (Niroomand, Hajilee, & Al Nasser, 2014).

Using a set of regression methodologies, Rahman et al. (2023) investigate the impact of trade openness on the cost of financial intermediation and bank performance. The results reveal that embedding higher trade openness reduces financial intermediation costs and improves bank performance (Rahman, Rahman, Rahman, & Masud, 2023). Using three sets of financial development indicators to evaluate efficiency, size, and competition as aspects of financial development, Zhang et al. (2015) examine the impact of trade and financial openness on China's financial development. Results from dynamic panel estimation methods indicate that financial openness and trade are significant determinants of financial efficiency and competition, while openness has a negative impact on the degree of financial development. The results further indicate that the marginal effect of openness on financial efficiency and competition is positive for more open provinces (Zhang, Zhu, & Lu, 2015).

*2.2.1 The Impact of Trade Openness on the Turkish Banking Industry*

The positive relationship between economic growth and trade openness is well-documented, with increased trade leading to higher productivity, innovation, and more efficient resource allocation. However, the impact of trade openness on the banking sector, particularly in developing economies such as Turkey, warrants further investigation.

In this study, we hypothesize that increased trade openness has a positive impact on the operational efficiency and growth of Turkish banks. This hypothesis is predicated on the unique characteristics of the Turkish financial system, where banks dominate the provision of financial services, and the corporate debt market (i.e., the bond market) remains significantly underdeveloped. Consequently, banks play a pivotal role in facilitating international trade transactions and funding the increased financial needs arising from trade activities.

In Turkey, banks are the primary financial intermediaries facilitating international trade transactions. As a result of the preeminent role the banking sector plays in providing financial services, export and import activities necessitate banking services such as trade finance, letters of credit, foreign exchange transactions, and payment processing. As trade volumes increase, the demand for these banking services rises, thereby enhancing operational efficiency through economies of scale and increasing revenue streams for banks.

The underdevelopment of the corporate debt market in Turkey also implies that private firms rely heavily on bank financing for their funding needs. As firms engage more in international trade, their financing requirements increase to support expanded production, improved inventory management, and investment in new technologies. Turkish banks, therefore, experience growth in their loan portfolios, driven by higher demand for credit from businesses involved in trade activities. This dynamic not only boosts the growth of banks but also encourages them to innovate and enhance their service offerings to meet the evolving needs of their clients.

*2.2.2. The Impact of Financial Openness on the Turkish Banking Industry*

We also hypothesize that financial openness has a positive influence on the efficiency and growth of Turkish banks. Given the dominance of banks in Turkey's financial services industry, the flow of monetary funds between Turkey and the rest of the world necessitates substantial banking intermediation.

Financial openness facilitates the cross-border movement of capital and finance, including the purchase of foreign assets by Turkish residents and the acquisition of domestic assets by non-residents, as well as borrowing from foreign entities and repaying external debt. The banking sector, as the primary intermediary in financial transactions, plays a crucial role in managing these capital and financial flows. Increased financial openness is expected to enhance the efficiency and growth of banks through several channels. This intermediation role increases the volume of banking transactions, resulting in higher fee income and improved economies of scale. The influx of foreign capital and finance can also enhance the funding base of banks, allowing for greater credit expansion and supporting their growth. By expanding its range of services to include international payment processing, foreign exchange operations, and trade finance, the Turkish banking sector as a whole is likely to achieve better risk management through increased diversification of services and exposure to international financial markets. An additional benefit of exposure to international financial markets is the incentivization of Turkish banks to improve their operational efficiencies and adopt advanced technologies and practices.

*2.3 Operational Efficiency and Growth*

Aftab et al. (2015) utilize CAMEL parameters to examine the impact of government versus private ownership on the performance of the banking industry. The results indicate that profitability is positively related to asset quality and management and negatively correlated with capital adequacy and liquidity ((Aftab, Samad, & Husain, 2015; Avkiran, 2011; Bongini, Laeven, & Majnoni, 2002; Calice, 2014; Hwa, Kapinos, & Ramirez, 2018; Jemric & Vujcic, 2002; Mayes & Stremmel, 2012; Nuxoll, O'Keefe, & Samolyk, 2003)2015). Calice (2014) proposes an early warning system for predicting bank insolvency based on a multivariate logistic regression framework, showing that the CAMEL indicators are significant predictors of bank insolvency (Calice, 2014). To predict bank failures and draw inferences about the stability of contributing bank characteristics, Mayes and Stremmel (2012) examine bank distress by incorporating CAMELS indicators that consider the bank-specific variables and macroeconomic conditions (Mayes & Stremmel, 2012).

Jemric and Vujcic (2002) conducted a data envelopment analysis to assess the relative performance of banks in terms of size, ownership structure, date of establishment, and asset quality compared to the market over a specified period (Jemric & Vujcic, 2002). Avkiran (2011) examines the relationship between bank DEA super-efficiency estimates and key financial ratios in informing pricing decisions and regulatory monitoring (Avkiran, 2011).

Bongini et al. (2002) investigate the performance of accounting data, credit ratings, and stock market prices as indicators of bank fragility. The differences in their ability to forecast financial distress are reported, leading to the policy conclusion that it is important to use a plurality of indicators simultaneously to assess bank fragility when information processing is costly (Bongini, Laeven, & Majnoni, 2002). Hwa et al. (2018) report that the effects of supervisory rating shocks on real economic activity are asymmetric, with bank downgrades leading to a pronounced decline while upgrades do not lead to an increase (Hwa, Kapinos, & Ramirez, 2018). Nuxoll et al. (2003) investigate the use of state-level economic data in statistical off-site monitoring models, concluding that it does not contribute to the models' ability to forecast bank failures and changes in the quality of bank assets. The results for the model predicting risky bank growth indicate that the inclusion of state-level economic data improves the model's predictive power (Nuxoll, O'Keefe, & Samolyk, 2003).

*2.4 Turkish Financial System Study*

Ararat et al. (2017) studied the corporate governance practices of Turkish public firms to build a corporate governance index predicting higher market value and firm-level profitability with firm random effects (Ararat, Black, & Yurtoglu, 2017; Ararat & Yurtoglu, 2021; Bildik, 2001; Demirkılıç, 2021; Ersan, Simsir, Simsek, & Hasan, 2021; Gemici, Gök, & Bouri, 2023; Kadırgan & Özlü, 2023; Stoupos, Nikas, & Kiohos, 2023; Ülkü & İkizlerli, 2012; Uz & Ketenci, 2008). Ararat and Yurtoglu (2021) studied the relationship between female representation on boards and firm value and profitability in Turkey (Ararat & Yurtoglu, 2021). Bildik (2001) investigates the intra-daily seasonalities of the stock returns in the Turkish stock market (Bildik, 2001).

Demirkılıç (2021) utilizes firm-level data on the composition and term structure of foreign currency assets and liabilities to investigate the balance sheet channels of depreciation for Turkish non-financial corporations (Demirkılıç, 2021). Ersan et al. (2021) argue that market reaction times to corporate announcements are slower, although markets react to positive news more quickly. Additionally, when frequent traders are more active in the market prior to announcements, the speed of price adjustments is slower (Ersan, Simsir, Simsek, & Hasan, 2021). Gemici et al. (2023) study the impacts of global and local factors on the risk appetite in Turkey, reporting stronger causal effects for regional factors such as CDS spreads and discussing policy implications (Gemici, Gök, & Bouri, 2023).

Kadırgan and Özlü (2023) examine the role of trade credit in the transmission of global liquidity conditions, utilizing data from both banks and firms in Turkey. Their study examines whether trade credit serves as a supplement to the reduction in bank lending when global liquidity becomes more constrained. The findings indicate that firms facing financial constraints receive less trade credit from suppliers and, in turn, reduce the amount of credit they offer to their customers, causing a ripple effect of tightening trade credit across supply chains. Stoupos et al. (2023) examine whether financial market behavior in the Turkish economy has forecasted the recent depreciation of the Turkish currency and claim that negative dynamics exist between the nominal exchange rate and the stock market index (Stoupos, Nikas, & Kiohos, 2023).

Ülkü and İkizlerli (2012) analyze the interaction between emerging stock returns and the trading of foreign investors, reporting that, contrary to the available literature, these investors are not uninformed positive feedback traders but can adjust their trading in line with the prevailing characteristics of the market (Ülkü & İkizlerli, 2012). Using panel unit root tests, Uz and Ketenci (2008) investigate the relationship between the nominal exchange rate and monetary variables, such as interest rates, prices, and monetary differentials, for Turkey and the new members of the European Union (Uz & Ketenci, 2008).

Turkey's relative poverty rate during our study period reflects a mixed trajectory within a narrow range, reaching its highest point of 8.9% in 2020 and its lowest point of 7.4% in 2017 and 2023. Holding all else equal, higher levels of poverty in Turkey are likely to impact the efficiency and growth of Turkish banks negatively. Higher poverty rates result in a smaller segment of the population being able to engage with the banking sector. Individuals in poverty are less likely to have disposable income for savings, investments, or purchasing financial products, resulting in reduced demand for banking services. Increased poverty also limits individuals' ability to save, resulting in lower deposit levels in banks. Since deposits are a key source of funding for banks, lower savings rates can constrain banks' ability to lend, invest, and grow.

In a high-poverty environment, the risk of loan default increases, as borrowers may struggle to repay due to unstable or insufficient income. This heightened credit risk can lead to higher non-performing loan ratios, which adversely affect the profitability and stability of banks. Moreover, high poverty levels can limit the incentives for banks to innovate and develop new financial products, as the potential market for such products is reduced. Without a substantial customer base, banks may be less willing to invest in innovation, which can stifle growth and efficiency (see Demirgüç-Kunt and Klapper (2013); Beck et al. (2009); Ghosh (2015); and Rajan and Zingales (2003) whose findings support our hypothesis).

Turkey's performance on the Global Innovation Index (GII) from 2010 to 2023, the period covered in our paper, shows a trajectory of gradual improvement. During the period from 2010 to 2015, Turkey was often ranked in the lower half of the GII, typically between 60th and 70th out of over 120 countries. The second half of the 2010s marked a phase of steady progress in Turkey's innovation indicators, with the country's GII ranking reaching the 50s range by 2020. By 2023, Turkey advanced further in the GII rankings, consistently placing in the 40s, which reflects the country's growing focus on innovation and strategic initiatives. We argue that innovation can play a critical positive role in enhancing the competitiveness, efficiency, and growth potential of the banking sector.

As the level of innovation in Turkey's economy rises, particularly in areas such as information and communication technology, Turkish banks are likely to adopt new technologies, processes, and business models that enhance their operational efficiency. Innovations in digital banking, automation, and data analytics can reduce transaction costs by streamlining operations, improving risk management, and enhancing customer service. For example, based on data from 32 countries over the period 1996–2010, Beck et al. (2016) demonstrate that the adoption of information and communication technologies and digital innovations in the banking sector results in significant cost reductions and efficiency gains.

Higher levels of innovation can also enable banks to diversify their product offerings and develop new financial services, such as mobile banking, online investment platforms, and fintech solutions. This diversification can attract new customers, increase market share, and drive revenue growth. In this regard, Berger et al. (2010) find that banks that innovate and diversify their products tend to experience higher growth rates.

## 3. Methodology

This section provides an overview of the research design and the rationale behind the methods utilized, presenting insights into the methodologies and models used to assess the impact of trade and financial openness on the operational efficiency and growth of Turkish banks, in the context of the selected variables. This study extends the conventional DEA model and utilizes CAMEL ratings to address efficiency evaluation challenges associated with inconsistent units of input or output variables. The banks studied in this study account for approximately 90% of the banking sector; thus, their data are representative of the banks operating in Turkey. Figure 1 depicts the framework of the two-stage CAMELG-DEA and Panel Data Analysis (PDA).

### 3.1 Data and Variables

A dataset comprising 18 banks operating in Turkey between 2010 and 2023 was compiled for this study using the financial statements of these banks available on the BRSA website (retrieved from https://www.bddk.org.tr). The variables related to individual banks used in the analysis were selected based on established banking literature. A detailed description of these variables is provided in Table 2.

**Table 2.** Variables

| Variable | Source | Type | Definition |
|---|---|---|---|
| Size | BRSA | Cat | Categorization of the banks on a 5-point scale, based on the number of branches they have in Türkiye. |
| Total Assets (C) | BRSA | Num | The sum of total assets |
| Shareholders' Equity(A) | BRSA | Num | Total assets minus total liabilities of the bank |
| Net Operating Profit/Loss(M) | BRSA | Num | Earnings of a bank from its primary operations |
| Total Comprehensive Income (E) | BRSA | Num | Change in the value of net assets from non-owner sources |
| Uncollectible Loans (L) | BRSA | Num | Receivables, loans, or other debts without a chance of being paid |
| Growth (G) | BRSA | Num | Growth in bank loans, assets, bank deposits, and other deposits |
| **PDA variable** | | | |
| Trade openness | WB | Num | (Exports + Imports) / GDP |
| Financial openness | WB | Num | (Net foreign assets + Net incurrence of liabilities)/GDP |
| Poverty index | TUIK | Num | Poverty rate (%) Risk of poverty: 40% |
| Innovation | GE | Num | GII index |

BRSA: Banking Regulation and Supervision Agency WB: World Bank TUIK: Turkish Statistical Institute GE: Globaleconomy.com

### 3.2 CAMELG-DEA Formulation

For each Turkish bank, following the concepts presented in the proposed Tone and Tsutsui (2010)This study proposes the formulations for the CAMELS-DEA model, which incorporates both desirable and undesirable variables. Table 3 provides the list of variables used in our CAMELG-DEA model.

**Table 3.** Notations of the CAMELG-DEA model

| | |
|---|---|
| $E_t^k$ | DMU efficiency associated with Turkish bank $k$ at term t, (t=1, ⋯, T) and (k = 1, ⋯, n) |
| $x_{i0t}^{k,g}$ | DMU input variables (desirable) of Turkish bank $k$ at term t, (t=1, ⋯, T) |
| $X_t^g$ | Matrix of $m_1 \times n$ dimension for $m_1$ input variables (desirable), $X_t^g = \sum_{k=1}^{n} x_{i0t}^{k,g}$ |
| $x_{i0t}^{k,b}$ | DMU input variables (undesirable) of Turkish bank $k$ at term t, (t=1, ⋯, T) |
| $X_t^b$ | Matrix of $m_2 \times n$ dimension for $m_2$ input variables (undesirable), $X_t^b = \sum_{k=1}^{n} x_{i0t}^{k,b}$ |
| $y_{r0t}^{k,g}$ | DMU output variables (desirable) of Turkish bank $k$ at term t, (t=1, ⋯, T) |
| $Y_t^g$ | Matrix of $p_1 \times n$ dimension for $p_1$ output variables (desirable), $Y_t^g = \sum_{k=1}^{n} y_{r0t}^{k,g}$ |
| $y_{r0t}^{k,b}$ | DMU output variables of (undesirable) Turkish bank $k$ at term t, (t=1, ⋯, T) |
| $Y_t^b$ | Matrix of $p_2 \times n$ dimension for $p_2$ output variables (desirable), $Y_t^b = \sum_{k=1}^{n} y_{r0t}^{k,b}$ |
| $\lambda_t^k$ | DMU inputs and outputs' intensive vector of Turkish bank $k$ at term t, (t=1, ⋯, T) |
| $l_t^k$ | DMU variable used in linear transformation of nonlinear DEA model of Turkish bank $k$ at term t, (t=1, ⋯, T) |

| $\Lambda_t^k$ | DMU intensive vector for link flows in Turkish bank $k$ at term t, where $\Lambda_t^k = l_t^k \lambda_t^k$ with the condition of continuity of link flows from term t to t + 1: $\sum_{k=1}^d \Lambda_t^k = \sum_{k=1}^d \Lambda_{t+1}^k$, (t=1,⋯, T-1) |

CAMELS-DEA model:

$$E_t^k = \min \frac{1}{1+\frac{1}{p_1+p_2}(\sum_{r=1}^{p_1} \frac{s_{it}^{k,g+}}{y_{rot}^{k,g}}+\sum_{r=1}^{p_2} \frac{s_{it}^{k,b+}}{y_{rot}^{k,b}})} \quad (1)$$

$$\text{s.t. } x_{r0t}^{k,g} l^k = X_t^g \Lambda_t^k + s_{it}^{k,g-} \quad (2)$$

$$x_{r0t}^{k,b} l^k = X_t^{,b} \Lambda_t^k - s_{it}^{k,b-} \quad (3)$$

$$y_{r0t}^{k,g} l^k = Y_t^g \Lambda_t^k - s_{it}^{k,g+} \quad (4)$$

$$y_{r0t}^{k,b} l^k = Y_t^b \Lambda_t^k + s_{it}^{k,b+} \quad (5)$$

$$s_{it}^{k,g-} \geq 0, s_{it}^{k,g+} \geq 0, s_{it}^{k,b+} \geq 0, s_{it}^{k,b-} \geq 0, \Lambda_t^k \geq 0, l_t^k > 0 \quad (6)$$

, where $s_{it}^{k,g-}, s_{it}^{k,g+}, s_{it}^{k,b+}, s_{it}^{k,b-}$ are slack variables on inputs and outputs in terms of desirable, and undesirable, respectively.

The CAMELG-DEA approach examines how to increase output without increasing input consumption, thereby enhancing overall output. This model reflects the efficiency of sustainable growth by selecting financial indicators that promote growth.

## 4. Empirical Results

This research employs a CAMELG-DEA method to assess efficiencies of Turkish banks. Table 4 and Table 5 provide descriptive statistics and a summary of CAMELG-DEA operational efficiency based on the CAMELG-DEA. Figure 2 illustrates the mean efficiencies of the Turkish banks. The findings are discussed within the discussion section.

- **Insert Table 4 Here-**
- **Insert Table 5 Here-**
- **Insert Figure 2 Here-**

Table 6 presents the significant relationship between the efficiency of Turkish banks and their type and size.

- **Insert Table 6 Here-**

Subsequently, to assess the relationship between bank efficiency and growth variables, random-effects (RE) and fixed-effects (FE) approaches are applied. Table 7 presents the results of the FE and RE models.

- **Insert Table 7 Here-**

The Hausman test is used to determine which model to choose (Hausman, 1978). Given that the *p*-value in Table 8 is smaller than 5%, the null hypothesis of the consistent random-effect model is rejected. Thus, the fixed-effect approach is used in analyzing the bank efficiencies.

- **Insert Table 8 Here-**

In Table 7, *trade openness, financial openness, poverty rate, and innovation* have significant relationships ($p$-value < 0.05) with bank efficiency in all models, except for the input-oriented model, where the poverty rate has a $p$-value > 0.05.

In addition to capturing cross-sectional variations across entities through static panel analysis, as reported in Tables 5-8, dynamic panel analysis is employed to mitigate potential endogeneity bias by incorporating lagged dependent variables as instruments In Table 9, the coefficients for *Trade Openness*, and *Innovation* exhibit positive values, indicating a positive association with the dependent variable, while the coefficient for *Financial Openness* and *Poverty Rate* are negative, suggesting an inverse relationship. The GMM results indicate statistically significant coefficients for the independent variables in the panel data, with a $p$-value of less than 0.05.

- **Insert Table 9 Here-**

The Sargan test returns a $p$-value below 0.05, indicating that the overidentifying restrictions are not violated when incorporating lagged dependent variables for estimation. This underscores the validity of the instruments and confirms the correct specification of the dynamic panel model. We address potential endogeneity bias by employing the Arellano-Bover/Blundell-Bond estimator, which includes lagged dependent variables (efficiency) as instruments. This correction is crucial for ensuring the consistency and unbiasedness of parameter estimates. Moreover, these lagged effects in the dynamic panel model indicate the persistence of relationships between variables over time.

The autocorrelation tests conducted with the Arellano-Bover/Blundell-Bond Estimator indicate no evidence of autocorrelation, as the $p$-value exceeds 0.05. Additionally, these results validate the Sargan test, confirming that the overidentifying restrictions are valid. The Wald test reports that a group of coefficients in the dynamic panel model is jointly significant, with a $p$-value of less than 0.05. This further underscores the validity and reliability of the dynamic panel model. These diagnostics provide valuable insights into the robustness of our model and the credibility of the estimated results.

5. Discussion

This section is organized around the analytical results concerning the empirical results obtained in this study as they relate to the growth variables.

### *5.1 Trade Openness*
The results from the output-oriented fixed-effects specification are shown in Table 7. The findings robustly support the hypothesis that there is a positive relationship between trade openness and the efficiency and growth of Turkish banks since the coefficient of TO has the predicted positive sign and is statistically significant at the 1% level. The results indicate that as Turkey's trade openness increases, so too does the demand for banking services associated with exports and imports, since these transactions predominantly go through the banking system due to the absence of a developed debt market for Turkish firms. This heightened demand has an expansionary impact on the scale and scope of banking activities, directly enhancing operational efficiencies and stimulating growth within the sector. Moreover, the increase in global trade volumes necessitates greater financing needs among businesses, further driven by the lack of alternative funding sources, such as a bond market. Turkish banks, therefore, find themselves at the core of financial intermediation for trade-related activities, benefiting from both increased transaction volumes and higher demand for credit. This alignment between trade openness and banking activity seems to have underpinned significant improvements in the operational efficiency and growth metrics of the banks, corroborating the theoretical predictions and illustrating the critical role of the banking sector in facilitating and capitalizing on Turkey's global economic engagements.

### *5.2 Financial Openness*

The coefficient on FO is also positive and statistically significant at the 5% level, confirming the hypothesis that financial openness significantly enhances the efficiency and growth of Turkish banks. The dominant role of banks in Turkey's financial sector positions them as key intermediaries in managing the flow of financial funds between Turkey and the international market. As financial openness increases, so does the influx of foreign capital and the cross-border movement of finance, which facilitates a broader range of financial transactions, from the purchase of foreign assets by residents to the acquisition of domestic assets by non-residents. This expansion in transaction volume notably increases fee income for banks and improves their economies of scale. Moreover, the additional capital inflows fortify the banks' funding bases, enabling greater credit expansion and underpinning substantial growth within the sector. The adoption of international payment processing, foreign exchange operations, and enhanced trade finance services also leads to improved risk management through service diversification and exposure to varied financial markets. Consequently, Turkish banks have been incentivized to adopt advanced technologies and practices, further driving operational efficiencies.

### *5.3 Innovation*

The estimated coefficient on IV has the expected positive sign; however, it is not statistically significant. The difficulty of accurately capturing the level of innovation within a country's economy and directly linking it to specific outcomes in the banking sector might be one explanation why the positive relationship between innovation in Turkey and the efficiency and growth of Turkish banks yielded a statistically insignificant result. It is also plausible that even as banks innovate, the level of competition within the banking sector and market saturation might limit the growth and efficiency gains from such innovations. For example, suppose most banks adopt similar technologies simultaneously. In that case, the competitive advantage and incremental gains from innovation may be diluted, leading to a statistically insignificant relationship in the aggregate data.

### *5.4 Poverty Rate*

Lastly, the coefficient of PR has the opposite of the expected negative sign and is statistically significant at the 5% level. One plausible explanation for the counterintuitive positive relationship between higher poverty levels and the efficiency and growth of Turkish banks is that a substantial segment of the population in high-poverty areas may be largely unbanked and not actively engaged with the formal banking sector. In this case, the data on banking efficiency and growth may not fully capture the interactions (or lack thereof) between the poorest segments and the banking sector. If the unbanked population is significant, any positive trends in banking metrics may reflect the sector's performance among a more affluent subset of the population, which may still be growing or utilizing banking services more efficiently, despite overall high poverty levels.

## 6. Conclusions

This study provides comprehensive insights into the impact of trade and financial openness on the operational efficiency and growth of Turkish banks. Our findings reveal that financial openness enhances growth by expanding the range of banking transactions and credit portfolios, driven by cross-border capital flows. Trade openness has a positive influence on banking efficiency and growth by increasing the demand for financial services associated with international trade activities. Meanwhile, the study also finds that higher poverty rates have a negative impact on bank performance, reducing access to financial services and limiting innovation within the banking sector. The research emphasizes the significance of policy frameworks that promote trade and financial liberalization while addressing the socioeconomic challenges associated with poverty. In doing so, the Turkish banking sector can further optimize its operational efficiency and maintain sustainable growth in a competitive global financial landscape. The results have significant implications for emerging economies seeking to balance openness with financial stability.

This research makes several significant contributions to the academic literature on trade and financial openness, as well as their impact on the banking sector, particularly in the context of an emerging economy like Türkiye. First, it extends the existing research by focusing on the combined effects of trade and financial openness on operational efficiency and growth in the banking sector, which has been less explored in previous studies. Utilizing the CAMELG-DEA model, this study introduces a novel approach to evaluating the performance of Turkish banks by integrating both desirable and undesirable variables in efficiency measurement. Second, by applying dynamic panel data techniques, the study addresses endogeneity concerns and provides robust empirical evidence on the long-term effects of openness on bank performance. Third, the findings provide policymakers with practical insights, highlighting the importance of promoting trade and financial openness while taking into account socioeconomic factors, such as poverty, which can hinder banking sector efficiency. These contributions provide a framework for further research in other emerging markets and offer actionable strategies for enhancing bank growth and efficiency in a globalized economy.

**Tables**

**Table 4.** Descriptive Results

|  | Operational Efficiency |
| --- | --- |
| Mean | 0.2214 |
| Median | 0.1183 |
| Standard Dev. | 0.2886 |
| Sample Variance | 0.0833 |
| Kurtosis | 3.1065 |
| Skewness | 2.1480 |
| Minimum | 1.64E-08 |
| Maximum | 1 |

Note: Operational efficiency examines how to increase output without increasing input consumption, thereby enhancing overall productivity.

**Table 5.** Statistical Summary of CAMELG-DEA operational efficiency

| Year | Mean | Median | Standard Dv. | Sample Var. | Kurtosis | Skewness | Min. | Max. |
| --- | --- | --- | --- | --- | --- | --- | --- | --- |
| 2010 | 0.1812 | 0.0779 | 0.3009 | 0.0905 | 5.6182 | 2.6053 | 0.0004 | 1.0000 |
| 2011 | 0.3559 | 0.1435 | 0.4152 | 0.1724 | -0.9923 | 1.0152 | 0.0238 | 1.0000 |
| 2012 | 0.1989 | 0.1339 | 0.2242 | 0.0502 | 10.2130 | 2.9591 | 0.0009 | 1.0000 |
| 2013 | 0.2591 | 0.1480 | 0.3024 | 0.0914 | 2.7601 | 1.9356 | 0.0103 | 1.0000 |
| 2014 | 0.1557 | 0.1464 | 0.1233 | 0.0152 | 6.4480 | 2.1943 | 0.0005 | 0.5578 |
| 2015 | 0.2428 | 0.1305 | 0.2847 | 0.0810 | 4.8505 | 2.3977 | 0.0441 | 1.0000 |
| 2016 | 0.1153 | 0.1232 | 0.0436 | 0.0019 | 1.4439 | 0.2671 | 0.0290 | 0.2235 |
| 2017 | 0.1025 | 0.0883 | 0.0599 | 0.0036 | 10.7061 | 2.9052 | 0.0298 | 0.3178 |
| 2018 | 0.1947 | 0.1228 | 0.2327 | 0.0541 | 8.9080 | 2.8568 | 0.0186 | 1.0000 |
| 2019 | 0.2851 | 0.1274 | 0.3374 | 0.1138 | 1.5603 | 1.7537 | 0.0216 | 1.0000 |
| 2020 | 0.1970 | 0.1043 | 0.2977 | 0.0886 | 5.3129 | 2.5220 | 0.0210 | 1.0000 |
| 2021 | 0.3577 | 0.1233 | 0.4135 | 0.1710 | -0.9910 | 1.0258 | 0.0173 | 1.0000 |
| 2022 | 0.2324 | 0.0937 | 0.3576 | 0.1279 | 1.8024 | 1.8536 | 0.0000 | 1.0000 |

**Table 6.** Bank efficiency using ANOVA on bank type and size

| Group | DF | Sum Sq | Mean of SQ | F-value | Pr(>F) |
| --- | --- | --- | --- | --- | --- |
| Bank_type | 3 | 0.357 | 0.11893 | 4.726 | 0.00322*** |
| Bank_size | 4 | 0.705 | 0.17623 | 7.004 | 2.5E-05*** |
| Bank_type:Bank_size | 3 | 5.611 | 0.00229 | 0.091 | 0.96499 |

Note: (.p-value<=0.1, * p-value<=0.05, ** p-value<=0.01, *** p-value<=0.005).

Table 7. Outputs of bank efficiencies from random-effect and fixed-effect models

| Hypothesis | Variables | FE | RE |
|---|---|---|---|
| | Cons. | 0.4608 | 0.0186* |
| | TO | 0.0062** | 0.0363* |
| | FO | 0.0286* | 0.1968 |
| | PR | 0.0191* | 0.0385* |
| | IV | 0.9385 | 0.2443 |
| | Obs. | 216 | 216 |
| | R-square | 0.0453 | 0.0264 |
| | Residual sum of squares | 21.604 | 11.825 |

Note: (.p-value<=0.1, * p-value<=0.05, ** p-value<=0.01, ***, p-value<=0.005); TO: Trade openness; FO: financial openness; PR: poverty rate; IV: Innovation

Table 8. Hausman test between the random and fixed methods

| |
|---|
| data: Efficiency ~ TO + FO + PR + IV |
| chisq = 12.849, df = 4, p-value = 0.0120 |
| alternative hypothesis: one model is inconsistent |

TO: Trade openness; FO: financial openness; PR: poverty rate; IV: Innovation

Table 9. The results of dynamic panel data using the Arellano-Bover/Blundell-Bond estimator as Generalized Method of Moments (GMM)

Number of Observations Used: 414

Residuals:

| Min. | 1st Qu. | Median | Mean | 3rd Qu. | Max. |
|---|---|---|---|---|---|
| -1.3133 | -0.1055 | -0.0279 | -0.0045 | 0.0416 | 1.1037 |

Coefficients

| | Estimate | Std. Error | z-value | Pr(>|z|) |
|---|---|---|---|---|
| Efficiency | 0.3614 | 0.0887 | 4.0713 | 4.676e-05*** |
| TO | 0.0317 | 0.0128 | 2.4788 | 0.0131* |
| FO | -0.0182 | 0.0060 | -3.0024 | 0.0026** |
| PR | 0.0439 | 0.0228 | 1.9243 | 0.0543. |
| IV | -0.0177 | 0.0077 | -2.2835 | 0.0224* |

Signif. Codes: 0 '***' 0.001 '**' 0.01 '*' 0.05 '.' 0.1 ' ' 1

| | |
|---|---|
| Sargan test: chisq (17) = 17.5625 | (p-value = 0.4169) |
| Autocorrelation test (1): normal= -2.9282 | (p-value= 0.034) |
| Autocorrelation test (2: normal= 1.5640 | (p-value= 0.1178) |
| Wald test for coefficients: chisq (5) = 154.1743 | (p-value=< 2.22e-16) |

**Appendix**

Table A1. Descriptive analysis of CAMELG and economy-wide variables

| Variable | Mean | Standard Deviation | Kurtosis | Minimum | Maximum |
|---|---|---|---|---|---|
| Total Assets | 2.02E+08 | 3.21E+08 | 1.54E+01 | 8.80E+04 | 2.90E+09 |
| Shareholders' Equity | 1.94E+07 | 3.03E+07 | 1.62E+01 | 1.57E+03 | 2.78E+08 |
| Net Op. Profit/Loss | 2.68E+06 | 7.16E+06 | 5.88E+01 | -9.76E+05 | 8.03E+07 |
| Total Comprehensive Income | 2.55E+06 | 8.38E+06 | 6.99E+01 | -6.27E+06 | 1.15E+08 |
| Uncollectible Loans | 2.39E+06 | 3.38E+06 | 2.76E+00 | 0.00E+00 | 1.52E+07 |
| Growth-Loan | -1.64E+07 | 3.94E+07 | 4.75E+01 | -5.33E+08 | 8.90E+06 |
| Growth-Asset | -2.16E+06 | 6.80E+06 | 5.73E+01 | -8.33E+07 | 1.36E+07 |
| Growth-Bank Deposit | 9.12E+05 | 4.62E+06 | 4.77E+01 | -3.82E+07 | 5.83E+07 |
| Growth-Other Deposit | 1.71E+07 | 5.14E+07 | 8.21E+01 | -1.33E+07 | 8.03E+08 |
| Financial and trade openness Variables (External Factors) | | | | | |
| Trade Openness | 15.322 | 4.0848 | 3.5477 | 12.082 | 27.0567 |
| Financial Openness | 3.8242 | 4.6414 | -1.0069 | -2.33 | 12.56 |
| Poverty Rate | 8.7391 | 0.8862 | -0.8623 | 7.4 | 10.31 |
| Innovation | 36.734 | 1.815 | -1.5637 | 34.1 | 39 |

**Table A2.** Conditional correlation for CAMELG variables (Lag2)

| | Total Assets | Shareholders' Equity | Net Op. Profit/Loss | Total Comp. Income | Uncollectible Loans | Growth |
|---|---|---|---|---|---|---|
| Total Assets | 1 | 0.9541 | 0.6289 | 0.6336 | 0.5963 | 0.3371 |
| Shareholders' Equity | 0.9541 | 1 | 0.6151 | 0.6074 | 0.6533 | 0.3317 |
| Net Op. Profit or Loss | 0.6289 | 0.6151 | 1 | 0.9854 | 0.1658 | 0.6469 |
| Total Comp. Income | 0.6336 | 0.6074 | 0.9854 | 1 | 0.1345 | 0.6034 |
| Uncollectible Loans | 0.5963 | 0.6533 | 0.1658 | 0.1345 | 1 | -0.0587 |
| Growth | 0.3371 | 0.3317 | 0.6469 | 0.6034 | -0.0587 | 1 |

**Table A3.** Stationarity test for CAMELG variables

| | ADF Statistic | Lags Used | KPSS Statistic |
|---|---|---|---|
| Total Assets | 9.9119 | 0 | 0.9377 |
| Shareholders' Equity | 5.1976 | 1 | 0.9490 |
| Net Op. Profit or Loss | 0.0788 | 7 | 0.6042 |
| Total Comprehensive Income | 0.5959 | 11 | 0.5379 |
| Uncollectible Loans | 0.8353 | 1 | 1.0415 |
| Growth | 2.3690 | 11 | 0.6103 |

**Table A4.** Multicollinearity test (VIF) for variables in panel regression analysis

| *Regression Statistics* | | | | | |
|---|---|---|---|---|---|
| Multiple R | 0.1342 | | | | |
| R Square | 0.0180 | | | | |
| Adjusted R Square | 0.0138 | | | | |
| Standard Error | 0.1685 | | | | |
| Observations | 954 | | | | |
| VIF | 1.0183 | | | | |
| ANOVA | | | | | |
| | df | SS | MS | F | *Significance F* |
| Regression | 4 | 0,494979 | 0,123745 | 4,354524 | 0,001706 |

| | | | |
|---|---|---|---|
| Residual | 949 | 26,96822 | 0,028418 |
| Total | 953 | 27,4632 | |

**Table A5.** Stationarity test for variables in panel regression analysis

| | ADF Statistic | Lags Used | KPSS Statistic |
|---|---|---|---|
| Efficiency | -6.8141 | 0 | 0.4978 |
| TO | -4.0476 | 11 | 0.7475 |
| FO | -1.8955 | 0 | 0.5968 |
| PR | -0.7311 | 0 | 0.8238 |
| IV | -1.2600 | 0 | 0.5368 |

**Table A6.** Heteroskedasticity test (Breusch-Pagan Test) for variables in panel regression analysis

| Lagrange multiplier statistic | p-value | f-value | f p-value |
|---|---|---|---|
| 5.0861 | 0.2785 | 1.2738 | 0.2933 |

**Table A7.** Mutual Information for variables in panel regression analysis

| | Efficiency | TO | FO | PR | IV |
|---|---|---|---|---|---|
| Efficiency | | 0.086 | 0.074 | 0.100 | 0.066 |
| TO | 0.093 | | 2.690 | 2.681 | 2.570 |
| FO | 0.088 | 2.681 | | 2.681 | 2.553 |
| PR | 0.110 | 2.675 | 2.681 | | 2.570 |
| IV | 0.055 | 2.581 | 2.581 | 2.581 | |

Note: There is no nonlinear relationship between Efficiency and the four variables (TO, FO, PR, and IV) when the mutual efficiency is less than 0.5. There are nonlinear relationships among TO, FO, PR, and IV when the mutual efficiency is larger than 1.

**Table A8.** OLS Regression Results

| | | PDA | | |
|---|---|---|---|---|
| Dep. Variable: | Efficiency | R-squared: | 0.122 | |
| Model: | OLS | Adj. R-squared: | 0.049 | |
| Method: | Least Squares | F-statistic: | 1.673 | |
| Prob (F-statistic): | 0.172 | Log-Likelihood: | 89.024 | |
| No. Observations: | 53 | AIC: | -168.0 | |
| Df Residuals: | 48 | BIC: | -158.2 | |
| Df Model: | 4 | Covariance Type: | non-robust | |

| | coefficient | std err | t | P>|t| | [0.025 | 0.975] |
|---|---|---|---|---|---|---|
| Intercept | -0.3538 | 0.401 | -0.882 | 0.382 | -1.160 | 0.452 |
| TO | 0.0022 | 0.004 | 0.620 | 0.538 | -0.005 | 0.009 |
| FO | 0.0016 | 0.003 | 0.594 | 0.555 | -0.004 | 0.007 |
| PR | 0.0090 | 0.017 | 0.547 | 0.587 | -0.024 | 0.042 |
| IV | 0.0080 | 0.007 | 1.100 | 0.277 | -0.007 | 0.023 |

| | | | | |
|---|---|---|---|---|
| Omnibus: | 17.777 | Durbin-Watson: | 1.938 | |
| Prob(Omnibus): | 0.000 | Jarque-Bera (JB): | 21.448 | |
| Skew: | 1.346 | Prob(JB): | 2.20e-05 | |

| | | | |
|---|---|---|---|
| Kurtosis: | 4.569 | Cond. No. | 2.53e+03 |

[1] Standard Errors assume that the covariance matrix of the errors is correctly specified.
[2] The condition number is large, 2.53e+03. This may indicate the presence of strong multicollinearity or other numerical issues.

**Figures**

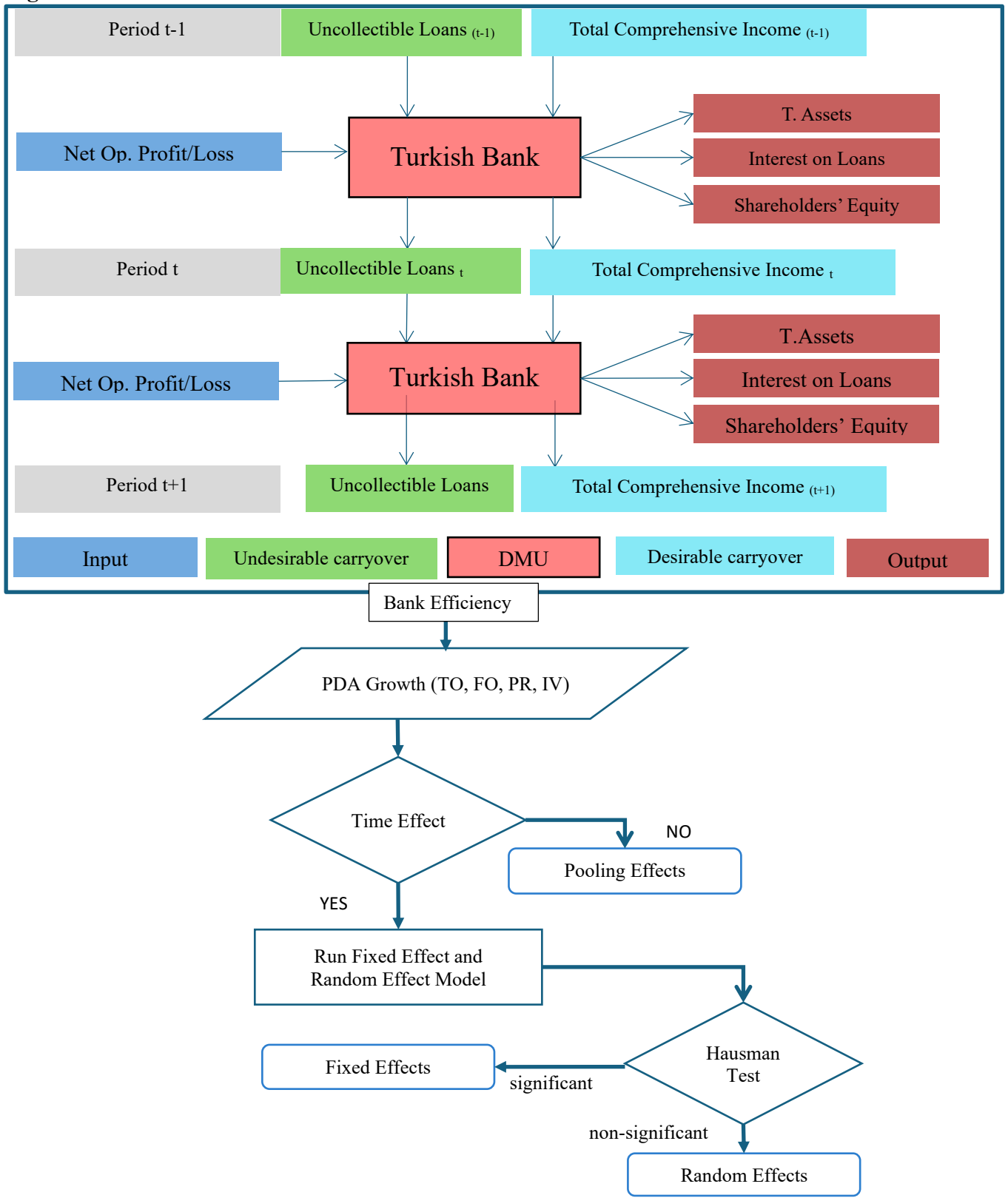

**Figure 1.** Framework of CAMELG-DEA and Panel Data Analyses

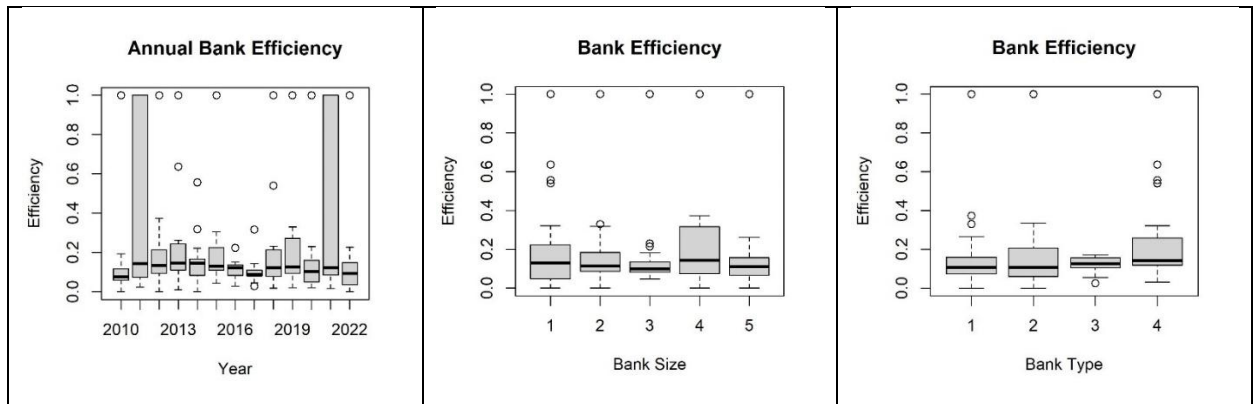

**Figure 2.** Bank Efficiency